# The Impact of Information Technology in Nigeria's Banking Industry

[1]Oluwagbemi Oluwatolani [2]Abah Joshua and [3]Achimugu Philip

**Abstract-**Today, information technology (IT) has become a key element in economic development and a backbone of knowledge-based economies in terms of operations, quality delivery of services and productivity of services. Therefore, taking advantage of information technologies (IT) is an increasing challenge for developing countries. There is now growing evidence that Knowledge-driven innovation is a decisive factor in the competitiveness of nations, industries, organizations and firms. Organizations like the banking sector have benefited substantially from e-banking, which is one among the IT applications for strengthening the competitiveness. This paper presents the current trend in the application of IT in the banking industries in Nigeria and gives an insight into how quality banking has been enhanced via IT. The paper further reveals that the deployment of IT facilities in the Nigerian Banking industry has brought about fundamental changes in the content and quality of banking business in the country. This analysis and clarification of how Nigerian Banks have used IT to reengineer their operations is detailed through literature review and observation. Three categories of variables that relate to the use and implementation of information technology devices were considered in this paper. These include the nature and degree of adoption of innovative technologies; degree of utilization of the identified technologies; and the impact of the adoption of IT devices on the bank operations.

——————————— ◆ ———————————

## 1 Introduction

The Internet is globally widespread in use, becoming an integral of IT within businesses as well as many homes [5]. A vast market has developed on the Internet, online purchasing and banking have been by-products of this growth. Many businesses have been quick to recognize and exploit the niche. The range product on-line is virtually inexhaustible and puts the Internet at the top of the list of convenience good, alongside ready-made meals. In this rapidly evolving modern society of which we are all a part, convenience has become crucial to survive the ever increasing pace of life.

In particular, e-business, one of the IT applications with the highest impact upon the global economy, is creating a new business environment [1]. As a growing number of companies launch new Internet-based business lines, many of the new technology advances occur as a result of their using the Internet to improve business processes. This often involves using the Internet to carry out business transactions. E-business has revolutionized the business sector in a way unprecedented in past centuries. It has fostered a new set of economic and social relationships. A critical use of the Internet is to develop and experiment with new business models. It is not technology by itself that makes or breaks an Internet venture, but the underlying innovation and adequacy of the adopted business approach. IT and e-banking have now become the key elements for strengthening the competitiveness of the national economy and improving the productivity and efficiency of both private and government banks. However, access to and use of these technologies remains extremely uneven [2].

Less developed economies are being left behind in the expansion of a global economy where knowledge is a key factor driving productivity growth. IT and e-banking contribute to the future of developing countries, Nigeria inclusive; underestimating their importance may ultimately increase the gap with industrialized countries. Most banks in the country look towards opportunities arising from the new marketplace. They also hope to benefit from the more pervasive and enduing effects of e-banking upon their

————————————————

1. *Oluwagbemi Oluwatolani is with the Department of Computer Science of Lead City University, Ibadan, Nigeria.*
2. *Abah Joshua is with the Computer Engineering Department of the University of Maiduguri, Maiduguri, Nigeria.*
3. *Achimugu Philip is with the Department of Computer Science of Lead City University, Ibadan, Nigeria.*

business organizations. They are adopting Internet-based technologies to craft lean production systems and improve their distribution efficiency. In this way, the competitiveness of banks can be greatly enhanced.

Furthermore banks have to provide an excellent service to customers who are sophisticated and will not accept less than above average service. Thus, the issue of service marketing in general, and banking services in particular has become one of the most important and modern directions which have witnessed a substantial expansion during the last years in almost all societies [2,3]. This is because the increasingly significant role which banking services have with the widening and variety that these services are characterized with, thus banking services have touched most aspects of contemporary societies life and activities.

## 2   The Nigerian Banking Industry

The structure of the Nigerian financial services industry changed drastically during the period under review, bringing about significant changes in the market [9]. Within the context of current developments and with increased breadth and depth of competition, the task of identifying the unique characteristics that will enable any bank outperform its peers is becoming more challenging. The industry is now characterized by the following interesting dynamics:

### 2.1   Developing Business Models

Nigerian banks are rapidly internationalizing; a trend most visibly demonstrated by the number of Nigerian banks opening branches across West Africa whilst new players, especially foreign banks may soon emerge. Many banks have returned to the capital market to shore up their shareholders' funds beyond the required minimum level, to enable them play more actively in the international arena.

### 2.2   Customer Sophistication

The gradual re-emergence of the Nigerian middle class has given rise to a class of knowledgeable and financially savvy customers. Their benchmarks for service quality have also risen, aided by the intense competition among financial service providers to attract new customers. It is no longer just sufficient to provide products, but to align these closely with specific customer segments and their identified expectations.

### 2.3   Technology

In response to the demands for quick, efficient and reliable services, industry players are increasingly deploying technology as a means of generating insights into customers' behavioural patterns and preferences. Well developed outsourcing support functions (technology and operations) are increasingly being used to provide services and manage costs (e.g. Automated Teller Machine networks, Cards processing, Bill presentment and Payments, Software Development, Call centre operations and Network management).

### 2.4   Regulation and Supervision

Regulators are also moving towards global best practices, as they gain a visibly improved appreciation of Basle II plus Compliance. The larger and more complex the bank, the greater the range of risks it faces, which is why at United Bank for Africa (UBA) for instance, have adopted self-regulatory methods by addressing risks through a rigorous enterprise-wide risk management framework.

However, the scope and dimension of financial services in the foreseeable future will be different from the present, in terms of the character of players, dynamism of business models, competitiveness, customer's expectations, the degree of internationalization, adjustment to technological trends and innovations, as well as the standards of the underlying infrastructure. UBA have positioned themselves in line with these emerging trends. Specifically, the bank is looking beyond Ghana (the most popular destination for most Nigerian banks right now), and consider other virgin territories in sub-Saharan Africa which hold great potential, in view of the expected inflow of donor reconstruction funds, oil exploration funds and increased regional trade [4]. For sometime now, the bank's management has been re-inventing the institution as a dynamic, people driven, customer-focused institution and above all, as a place where customers are not just happy to bank, but employees (including out-sourced staff ) are happy to work with adequate provision and implementation of the current information technology support systems.

## 3 Electronic Banking and its Revolution in Nigeria

Electronic banking can be described as the act of carrying out the business transaction of a bank using electronic devices. Examples of electronic devices that are used include Computer Systems, Global System for Mobile Communication (GSM) phones, Automated Teller machine (ATM), Internet facilities, Optical Character Recognition (OCR), Smart Cards, etc. E-banking is about using the infrastructure of the digital age to create opportunities, both local and global. E-banking enables the dramatic lowering of transaction costs, and the creation of new types of banking opportunities that address the barriers of time and distance. Banking opportunities are local, global and immediate in e-banking [6, 7].

The evolution of e-banking dates back to 1986 when the banking sector in Nigeria was deregulated [10]. The result of this deregulation brought far-reaching transformation through computerization and improved bank service delivery. Competition with new products became keen within the system while customer sophistication posed a challenge for them, hence the reengineering of processing techniques of business activities encourage the automation of financial services especially among new generation of commercial and merchant banks.

In effect, the emergence of a crop of new generation banks following the liberalization of bank licensing motivated the introduction of high technology in the Nigerian banking system.

Some of them considered the "arm-chair" brick and mortar approach to banking of the old generation banks as having no regard for the customers and therefore an identified weakness they can exploit on.

These new banks discovered that the evolving technology at the global level could be applied to greater advantage in the Nigerian financial landscape [8]. That indeed paid off for some of them, as customers, who ordinarily would have found it almost impossible to leave the banks they were already familiar with for a new one that was yet to find its feet, quickly noticed the difference in the available products and service delivery systems of the two categories of banks (old and new generation). The customers without hesitation opted to pay for the extra values that would satisfy the extra-personalized product services and the attendant personalized marketing.

They therefore demonstrated their preparedness to switch from one bank to the other, the old ties notwithstanding; as they identified gaps in the service delivery of their original banks [6]. E-banking makes use of certain identification features before access/permission is granted by the bank to its users. A widely used identification feature today is the use of personal identification number (PIN). These are usually a series of codes which is only known to the customers who owns the account or anyone else the person(s) wished to have access to his account. Permission to perform financial transactions is immediately granted by the banks once this PIN is quoted.

## 4 Impact of IT in Nigeria's Banking Industry

The following include some of the major impacts of information technology in Nigeria's banking system:

### 4.1 GSM Banking

This mode of e-banking makes use of the Global System for Mobile communication (GSM) phones as the primary electronic device. GSM has improved the operational efficiency of many banks in the country. The mobile banking services basically allow customers to operate their accounts with the operating banks from mobile phones to a large extent as long as their phones and network support SMS (short messaging service). The user could be able to check account balance up to his two last transactions.

### 4.2 Automated Teller Machines (ATMs)

ATMs are a computer-controlled device that dispenses cash, and may provide other services to customers who identify themselves with a Personal Identification Number. ATM dispenses cash at any time of the day and night, unlike the traditional method where customers have to queue for a very long time in order to withdraw cash or transfer funds.

### 4.3 Adoption of the ICT Integrated Project

Banks in Nigeria have successfully completed information and communication technology integration project which enables them to communicate easily across as many employees as possible within and outside the country to deliver radically-enhanced customer-centric services.

### 4.4 Funds Transfer

Customers can now electronically transfer funds across the globe without any problem or delay as compared to the traditional method before the advent of information technology when funds are seriously delayed before they are delivered to the recipients.

### 4.5 On-Line Banking

With the aid of information technology, online banking provides the opportunity of paying bills and performing transactions of any kind electronically. Electronic payments can be credited or debited the same day. Customers can make payments for goods or services without necessarily coming in contact with physical cash and running the risk of handling a large amount of money.

### 4.6 Electronic Mail

Information technology has given rise to electronic mail which improves communication between individuals, external parties and the bank within or across various geographical regions or boundaries. The availability of online information provides bankers and customers with a powerful vehicle for research.

### 4.7 Bankers Automated Clearing Services

This involves the use of Magnetic Ink Character Reader (MICR) for cheque processing. It is capable of encoding, reading and sorting cheques. Also, request for cheque books or purchase of draft can be made and granted via electronic devices that are web-enabled.

Summarily, the impact of information technology in banking industries in Nigeria cannot be over-emphasized. It has provided flexible and convenient services to customers. Most current e-banking applications make use of the Internet which allows customers to obtain current account balances at any time. Customers do not need to bother themselves once money have been deposited or withdrawn from their accounts as most banks in Nigeria employs the use of short message service (SMS) to intimate customers of their balances immediately the transaction is performed.

## 5 Conclusion

There are indeed no doubts that majority of organizations including the banks have taken the advantage of IT to enhance their operations. Today most of them have website on the Internet in order to extend their services globally, provide executive services and promote quality of service delivery [8]. Driven by their ambitious aspirations to dominate the African financial services landscape, and under the leadership of a dynamic and visionary management team through information technology, Nigerian banks has been rapidly transformed from being just a bank to a one-stop-shop financial solutions provider. As the economies of Nigeria and Africa continues to improve, following the established path of other emerging markets; that is, increased political stability, improved government finances, growing domestic consumer demand, high commodity prices and significant improvement in other economic indicators, the banks in Nigeria are well positioned as a warrant on the African renaissance story.

It is expected that when the 3G network is operational, it will boost m-Commerce activities in Nigeria but may require further investment in the quality of cell phones. However, there are enormous opportunities for m-Commerce implementation in Nigeria based on the rate of growth and the diffusion of mobile devices. There is prospect for patronage but may be dependent on the available services.

## 6 References

[1] J. O. Adetayo, S.A. Sanni, and M.O. Ilori, "The act of Information Technology on Product Marketing: A Case Study of Multinational Company in Nigeria" Technovation, Elsevier Science Ltd, 1999.

[2] A.A. Agboola, "Impact of Electronic Banking on Customer Services in Lagos, Nigeria" in Ife Journal of Economics and Finance. Department of Economics, O.A.U, Ile-Ife, Nigeria, vol. 5, Nos. 1&2, 2000.

[3] A.A. Agboola, "Inform Technology, Bank Automation and Attitude of Workers in Nigerian Banks" in Journal of Social Sciences, Kamla-Raj Enterprises, Gali Bari Paharwali, India, 2003.

[4] S. A. Akpore, "The Backbone of Banks' Service Regeneration", Money watch, J 22, p23.1998.

[5] J.H. Boyett, and J. T. Boyett, "Beyond Workplace 2000: Essential Strategies for the New American Corporation, New York" Dutton, 1995.

[6] J, Hawkins, "The Banking Industry in the Emerging Market Economies: Competition, Consolidation and Systemic Stability: An Overview". BIS Papers No. 4, pp: 1-44, 2001.

[7] A, Jide, "Don't open an account, if it isn't an E-bank". Jidaw collective Available at http://www.Jidaw.com//article.html, 2002.

[8] L.O. Ugwu, T.O. Oyebisi, M.O. Ilori and E.R. Adagunodo, "Organizational Impact of Information Technology on Banking and Insurance Sector in Nigeria" TECHNOVATION Vol. 20, No 12, 2000.

[9] E.W. Woherem, "Information Technology in the Nigerian Banking Industry" Spectrum Ibadan, 2000.

[10] Stamoulis, D, "How Banks Fit in an Internet Commerce Business Activities Model". Journal of Internet Banking and Commerce, Vol. 5, No. 1.

**ACHIMUGU Philip** is a Lecturer at the Department of Computer Science of Lead City University, Ibadan, Nigeria. He holds B.Sc and M.Sc degrees in 2004 and 2009 respectively in Computer Science. He is also a PhD student at the Obafemi Awolowo University, Ile-Ife, Nigeria in the same field. He has over 18 publications in reputable journals and referred learned conferences both home and abroad. His research area is mainly software engineering with emphasis on: Development techniques, Development tools, Software products architecture and Usability.

**OLUWAGBEMI Oluwatolani** is a Lecturer at Lead City University, Ibadan, Nigeria. She can be reached at Department of Computer Science (Room 118), Faculty of Information Technology and Applied Science Building. She Holds B.Sc and M.Sc degrees in 2005 and 2011 respectively. Her research interests include computer-based information systems. She has published about 15 articles in locally and internationally reputable journals and learned conferences.

**Abah Joshua** is a Lecturer at the Department of Computer Engineering of the University of Maiduguri, Nigeria. He Holds B.Sc and M.Sc degrees in Computer Science in 2005 and 2011 respectively. His research interests include computer networks and communication. He has published many articles in locally and internationally recognized journals and learned conferences.